\documentclass[twocolumn,secnumarabic,amssymb, nobibnotes, aps, prl, reprint,longbibliography]{revtex4-1}
\pdfoutput=1

\usepackage{graphicx}
\usepackage{float}
\usepackage{subfig}
\usepackage{color}

\newcommand{\uts}{U}
\usepackage{amsmath}
\begin{document}

\title{The lower limit for time resolution in frequency modulation atomic force microscopy}%

\author{Zeno Schumacher}%
\email{zenos@physics.mcgill.ca}
\affiliation{Department of Physics, McGill University, Montreal, Quebec, Canada H3A 2T8}
\date{August 22, 2016}
\author{Andreas Spielhofer}
\affiliation{Department of Physics, McGill University, Montreal, Quebec, Canada H3A 2T8}

\author{Yoichi Miyahara}
\affiliation{Department of Physics, McGill University, Montreal, Quebec, Canada H3A 2T8}

\author{Peter Grutter}
\affiliation{Department of Physics, McGill University, Montreal, Quebec, Canada H3A 2T8}

\begin{abstract}
Atomic force microscopy (AFM) routinely achieves structural information in the sub-nm length scale. Measuring time resolved properties on this length scale to understand kinetics at the nm scale remains an elusive goal. We present a general analysis of the lower limit for time resolution in AFM. Our finding suggests the time resolution in AFM is ultimately limited by the well-known thermal limit of AFM and not as often proposed by the mechanical response time of the force sensing cantilever. We demonstrate a general pump-probe approach using the cantilever as a detector responding to the averaged signal. This method can be applied to any excitation signal such as electrical, thermal, magnetic or optical. Experimental implementation of this method allows us to measure a photocarrier decay time of $\sim$1~ps in low temperature grown GaAs using a cantilever with a resonance frequency of 280~kHz.
\end{abstract}

\maketitle

\section{Introduction}

Understanding structure-function relations at the nanometer length scale is important for many fundamental as well as applied questions in material science, engineering as well as biology. Since its invention more than 30 years ago, Atomic Force Microscopy (AFM) has demonstrated the capability of routinely determining structure with sub-nm resolution. Many interesting properties including surface potential \cite{Gross2014,Schuler2014}, electrical conductivity and density of states \cite{Park2002,Pingree2009,Roy-Gobeil2015} or mechanical parameters \cite{Gnecco2015,Bennewitz2005,Paul2014} can be simultaneously measured with AFM. Currently, most AFM measurements of structure-property relations are slow, on time scales of milliseconds-hours \cite{Schirmeisen2004,Coffey2006}. Much effort has been put into adding fast time-resolution to understand dynamics and kinetics on time scales of nano- to milliseconds \cite{Takihara2008,Coffey2006,Shao2014,Karatay2016}. A common assumption in AFM is that the slow time resolution of AFM is related to the mechanical oscillating probe with resonant frequencies of 0.1-1~MHz and slow detection and feedback electronics \cite{Ando2012,Ando2008a}. 

In 1990, R. Hamers and D. G. Cahill presented a first approach to ultrafast time resolution in scanning capacitance microscopy with pulsed laser illumination \cite{Hamers1990,HamersR.J.Cahill1991}. They pointed out that a time resolution faster than the bandwidth of the detection electronics can be achieved and is only limited by the time response of the underlying physical process of the sample. The proposed idea uses a stimulus $S(t)$  to generate a decaying response in the sample $R(S(t))$. For a linear responding system, the average signal will correspond to the average of the stimulus. However, for a non-linear response the average of the response $\langle R(S(t))\rangle$ will depend on the temporal distribution of the stimulating signal, $\langle R(S(t))\rangle \neq  R(\langle S(t)\rangle)$. We will extend this key idea in more general terms and thereby establish the lower limit for time resolution of AFM in terms of the minimal measurable energy for frequency modulation (FM) AFM.

In scanning tunneling microscopy fast time resolution in the picosecond range has been achieved by optical pump-probe excitation and non-linear mixing in the tunnel junction \cite{Yoshida2013,Yoshida2012,Terada2010b,Terada2010,Yoshida2014,Cocker2013,Grafstrom2002}. Note the important difference between the non-linear mixing in STM and the non-linear response to a stimulus mentioned above used in this study. STM is limited to conducting samples due to the need of a tunnel current, a limitation not existing for AFM. In addition, small perturbations of the tunnel gap as a result of synchronous thermal expansion is a cause of potential artifacts. Fast time resolution in AFM was achieved in 2015 by Jahng et al. to measure the dynamics in silicon naphtalocyganine using a technique named photo induced force microscopy (PIFM) \cite{Jahng2015}. PIFM measures the force generated by the light induced dipoles in the tip and sample. This indicates a great potential for extending AFM capabilities to measure fast processes. However, PIFM is limited to the near field dipole interaction between the tip and the sample induced by an optical pulse.

In the following, we will describe how to generically measure a signal decaying in time (i.e. a non-linear signal) with frequency modulation AFM (FM-AFM). This general analysis is not limited to a specific response or time constant and can, therefore, be applied to any timescales and any stimulating signal measured by a mechanical oscillator. It will also allow us to understand what determines the ultimate limitations of time resolution measurements by AFM.


\section{The Lower Limit for Time Resolution in AFM}

Most sub-nm structural determination by AFM is performed in the FM-AFM mode, where the resonance frequency shift of the oscillating cantilever is a measure of the tip-sample interaction \cite{Albrecht1991}. 
 
To measure the frequency shift of a cantilever one has to measure at least half a cycle of oscillation, $T_0/2$.  Consequently, one would assume that any changes faster than $T_0$ would not be resolvable in the frequency shift in real time. In the following, we will discuss how to overcome this apparent limitation. To generalize the discussion, we will apply our analysis to a generic time dependent potential energy $U(t)$ of the AFM cantilever. The change in $U(t)$ due to the time dependent tip-sample interaction $U_{\text{ts}}(t)$ will translate to a measurable frequency shift \cite{Durig1999,Giessibl1997,Garcia2002,Kantorovich2004}.

Following the approach of Hamers and Cahill \cite{Hamers1990}, an external signal $S(t)$ is applied to modulate the potential energy of the harmonic oscillator $U(t)$ through modulating $U_{ts}(t)$. In FM-AFM, such modulations are often sine waves, e.g. in Kelvin Probe Force Microscopy (KPFM), an AC voltage is applied to modulate the electrostatic $F_{ts}$.  In the following, we will apply a pulsed signal to modulate the tip-sample potential. The modulation signal can be of any nature including electrical, thermal, magnetic or optical, as long as it generates a modulation of the potential energy $\uts$ which translates to a modulation of the force and force gradient.

Assume two pulse trains which can be delayed in time with respect to each other. Additionally the second pulse train (probe pulse) is modulated to achieve a baseline/differential measurement. Each pulse will lead to an increase of the tip-sample interaction energy and a change in the energy of the harmonic oscillator $\uts$ followed by a characteristic decay time: 

\begin{equation}
\Delta \uts(S(t)) = \begin{cases}
 U_{0} & \text{~for~} 0 < t < T_{p} ,\\
U_{0} e^{-(t-T_p)/\tau} & \text{~for~} T_{p} < t < T_{rep} ,
\end{cases}
\end{equation}
with a pulse duration of $T_{p}$, a response amplitude $U_{0}$ and a repetition rate of $1/T_{rep}$. 

The decay time $\tau$ is the property of interest which we want to measure with AFM. In the following we will focus on the case when the decay and pulse width are shorter than the cantilever oscillation period of 1-10 $\mu$sec (and thus substantially shorter than the AFM electronic response time limited by the frequency detector, typically phase-locked loop (PLL),  whose  bandwidth is typically less than 1~ms).  Since the time constant $\tau$ can be recovered from the average signal, the cantilever itself can be used as an averaging tool. The only necessary requirement is that the cantilever is capable of resolving the change in potential energy between two different time delay settings. The lower limit for this is given by the minimal detectable energy as described by Albrecht et al.  \cite{Albrecht1991} and Smith \cite{Smith1995}.

The averaged potential energy over one cycle of $T_{rep}$ is calculated below. Considering only terms of the average potential energy which depend on the time delay we find the following expression:

\begin{equation}
\label{eq:avgTrepbaselinecontibution}
\langle \Delta \uts(S(t)) \rangle = \frac{ \tau U_0}{T_{rep} } (1-e^{-T_d/\tau} -e^{-(T_{rep}-T_d-2T_p)/\tau}) .
\end{equation}

The last term in equation \ref{eq:avgTrepbaselinecontibution} can be neglected by implementing a repetition time $T_{rep}$ that is much larger than the pulse width, the delay and the decay time. The remaining expression now allows us to determine the minimum measurable time constant $\tau$ in terms of the minimal detectable energy by AFM. Therefore, it is no longer a question of how fast the measurement can be done, but rather how small of a potential energy change can be measured. To determine this, we recall the well-known expression of the thermal limit of AFM \cite{Albrecht1991}. By equating the thermal noise dominated minimum measurable energy for on resonant detection in AFM to the change in potential energy due to the decay from equation \ref{eq:avgTrepbaselinecontibution}), we obtain \cite{Smith1995}:

\begin{equation}
\label{eq:limit_time_energy}
\frac{ \tau  U_0}{T_{rep} } (1-e^{-T_d/\tau}) = \frac{2 k_B T}{\pi f_0 Q \tau_s}
\end{equation}

with the resonance frequency $f_0$, the Q-factor of the cantilever $Q$, the Boltzmann constant, $k$ and temperature $T$. $\tau_s$ is the integration time (proportional to the inverse of the bandwidth) used in the experiment. 
Inspecting equation \ref{eq:limit_time_energy} we observe that our average potential energy and thus signal increases with higher repetition rate (i.e. shorter time between pulse pairs) since the system is pumped more often. The signal also depends on the initial response amplitude $U_0$. Thus, having a large response to the simulating signal is obviously beneficial. The faster the signal decays (smaller $\tau$), the smaller the averaged measured potential energy change. 

Assuming $T_d \ll \tau$, the minimum decay time constant is equal to:

\begin{equation}
\label{eq:limit:time:energy:10}
\tau  = \frac{T_{rep}}{U_{0} (1-e^{-\beta})}\frac{2 k_B T}{\pi f_0 Q \tau_s}
\end{equation}

with $\beta = T_{d}/\tau$, e.g. the ratio between the delay time and the decay time constant.

Note that we chose the simplification of having the same response amplitude for both the pump and probe pulses. This would be the case when the stimulation is saturating the signal, e.g. exciting all carriers in the probe volume. Further analysis can be found in the supplementary information, discussing a variable response amplitude dependent on the delay time, like ground state bleaching. The end result presented here, however, holds in all cases.

\section{Experimental Implementation}

\begin{figure}[ht]
\centering
\includegraphics[width=1\linewidth]{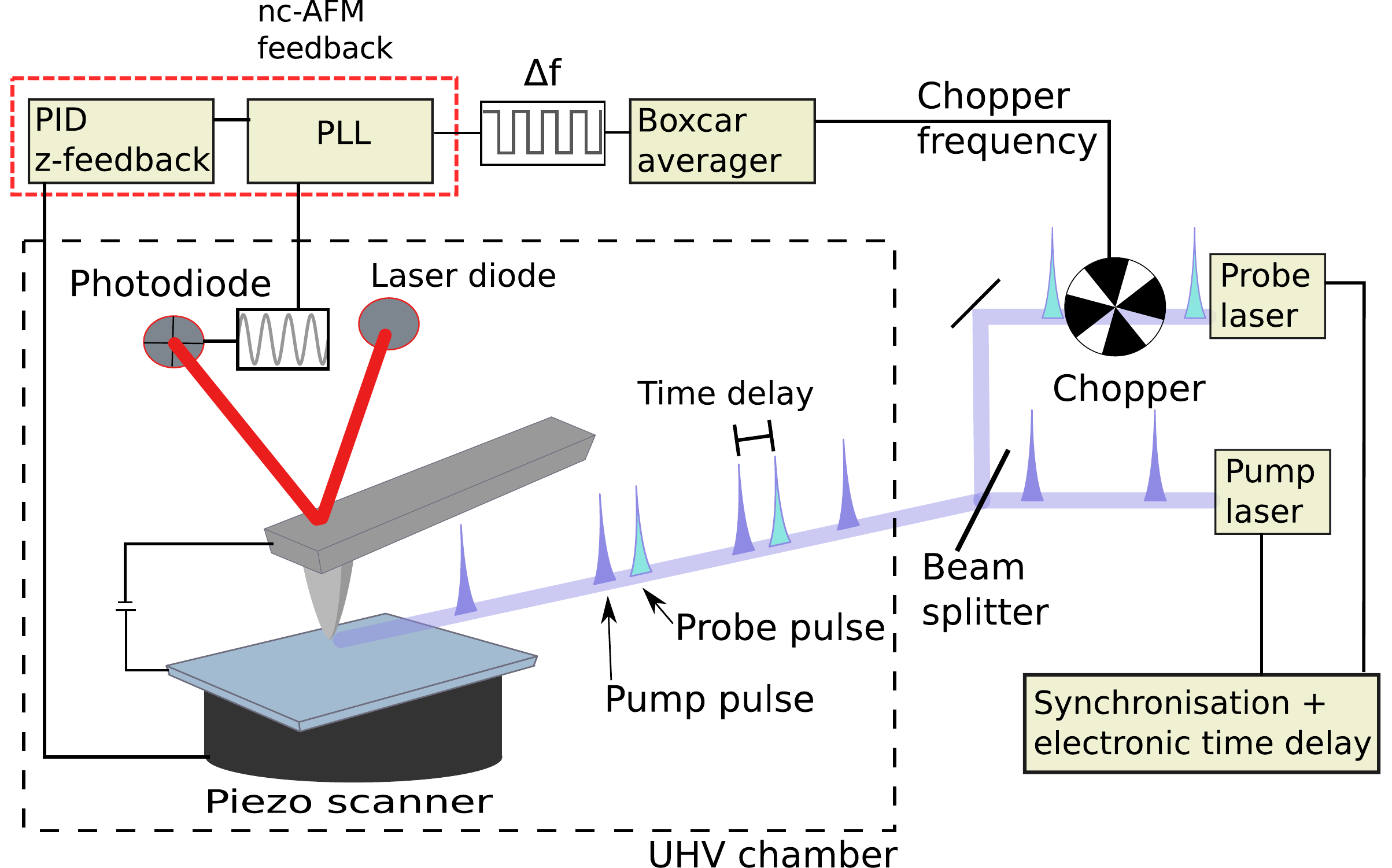}
\caption{A pump pulse train (blue) and a probe pulse train (green) are delayed with respect to each other. The probe pulse train is additionally modulated by a chopper. The final pulse train pattern consists of the delay pulse trains with and without the probe beam present. The sample and cantilever react to this pulse train resulting in a modulation of the frequency shift.}
\label{fig:modulationscheme}
\end{figure}

The proposed modulation scheme shown in figure \ref{fig:modulationscheme} is generic and can be implemented with an arbitrary pump and probe signal. Any stimulating signal leading to a decay, which can be applied as a pulse train, can be used. To determine decay times, two pulse trains are needed that can be delayed in time with respect to each other. 

We experimentally demonstrate the implementation and measurement by AFM of ultrafast photocarrier decay times in low temperature grown GaAs (LT-GaAs). LT-GaAs has a carrier lifetime in the low picosecond regime, depending on the growth conditions \cite{Gupta1991a}. During the LT growth process, many point defects are created \cite{Segschneider2002,Harmon1993}. These defects lead to a short diffusion length of the injected charge carrier and hence a short lifetime. For our measurement, we used a LT-GaAs sample (TeTechS Inc., Ontario, Canada) with a carrier lifetime of about 1 ps. A 1.5 $\mu$m LT-GaAs layer was grown on a 200~nm AlAs etch stop layer. The substrate consists of approximately 600 $\mu$m thick semi-insulating GaAs. The LT-GaAs layer is thicker than the penetration depth of the used illumination and therefore no contribution of the substrate is expected. The penetration depth was calculated to be 653~nm for an illumination with 780~nm and 253~nm for illumination with a 610~nm wavelength, respectively \cite{Jellison1992}. 

Our UHV AFM setup based on a JEOL 4500 SPM system was combined with an ultrafast pumped laser source. Commercial Pt coated cantilevers (PPP-NCHPt, Nanosensors) are used ($f_0 = 279647~\text{Hz}, Q = 11860$). An oscillation amplitude of 6~nm was typically used with the contact potential difference (CPD) compensated. The tip was lifted by 1~nm from a -2~Hz frequency shift setpoint with an applied tip bias of -950~mV with respect to the CPD between tip and sample. A 780~nm laser (Toptica FemtoFiber Pro NIR,80.1~MHz, 94 fs pulse width) is used with a power of 29.8~mW (0.182 nJ/cm$^2$) and a tunable visible wavelength laser (Toptica FemtoFiber Pro TVIS, 80.1~MHz, 0.4~ps) is adjusted to a 610~nm wavelength with an average power of 5.125~mW (0.032 nJ/cm$^2$). The fluence is low enough to avoid excessive heating of the AFM tip \cite{Milner2010}. The 610~nm beam is chopped at a low frequency (233~Hz) to generate a modulation of the frequency shift. Both lasers are locked to the same repetition rate with the variable repetition rate (VAR) and Laser repetition rate control (LRC) option from Toptica included in the NIR FemtoFiber Pro. A delay between the lasers can be adjusted electronically by adding a phase offset to the phase lock loop. This electronically controlled delay was used to delay the two laser pulses relative to each other. The RMS-jitter of the two locked lasers is measured to be 130~fs.

To measure the average signal generated by the delayed pulse trains, the probe pulse is chopped which translates into a modulation of the frequency shift. Here, the modulation of the frequency shift is measured with either a gated integration setup or a direct sideband lock-in detection \cite{Zerweck2005}. The chopping of the probe beam enables an accurate and drift-free measurement of the electrostatic tip-sample interaction and thus allows the measurement of the decay of photo-excited charges \cite{Schumacher2016}.

It is important to note that the chopping of the probe beam needs to be slow enough to be resolved by the PLL. Otherwise, the measured signal will either be influenced by the PLL dynamics or not reach the steady state signal. The frequency shift can then be integrated over a short time interval when no transient effects from the PLL are present. Having a boxcar integration with baseline subtraction allows for the direct measurement of the probe beam effect (see equation \ref{eq:avgTrepbaselinecontibution}). This difference will decay with the sought after characteristic time constant  $\tau$ while the delay between the two pulse trains is swept across the range of interest. 

A lock-in measurement at the probe pulse chopping frequency gives information about the difference between the two stages, hence the decay times. For imaging purposes, such a measurement might be well suited. However, if more information needs to be extracted from the change in signal, such as the change in capacitance gradient shown by time domain (TD)-KPFM \cite{Schumacher2016}, only a gated integration in the time domain will provide this additional information.

\begin{figure}[ht]
\centering
\includegraphics[width=1\linewidth]{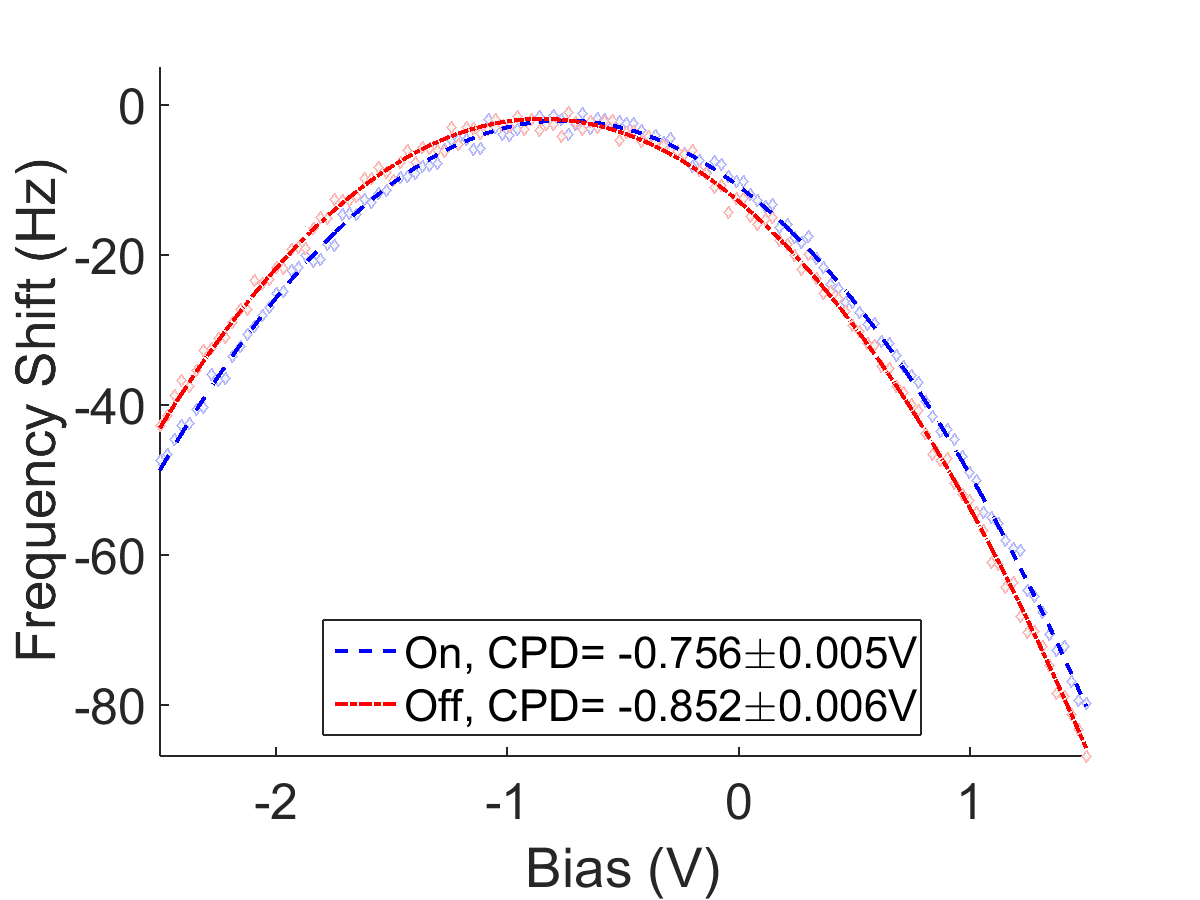}
\caption{Surface photovoltage of Lt-GaAs measured with TD-KPFM. A surface photovoltage of $96\pm8$~mV is measured.}
\label{fig:ltgaas:spectroscopy}
\end{figure}

A UHF lock-in amplifier from Zurich Instruments with the boxcar averaging option is used for data acquisition. All sideband lock-in measurements are performed with a HF2PLL from Zurich Instruments with direct sideband detection. To verify the change in CPD under pulsed illumination, the LT-GaAs sample is illuminated with 780 nm and TD-KPFM is used to measure the surface photovoltage under pulsed illumination \cite{Schumacher2016}. A surface photovoltage of  $96\pm8$~mV  is measured for this sample (see figure \ref{fig:ltgaas:spectroscopy}).

\begin{figure}[htp]

       \includegraphics[width=1\linewidth]{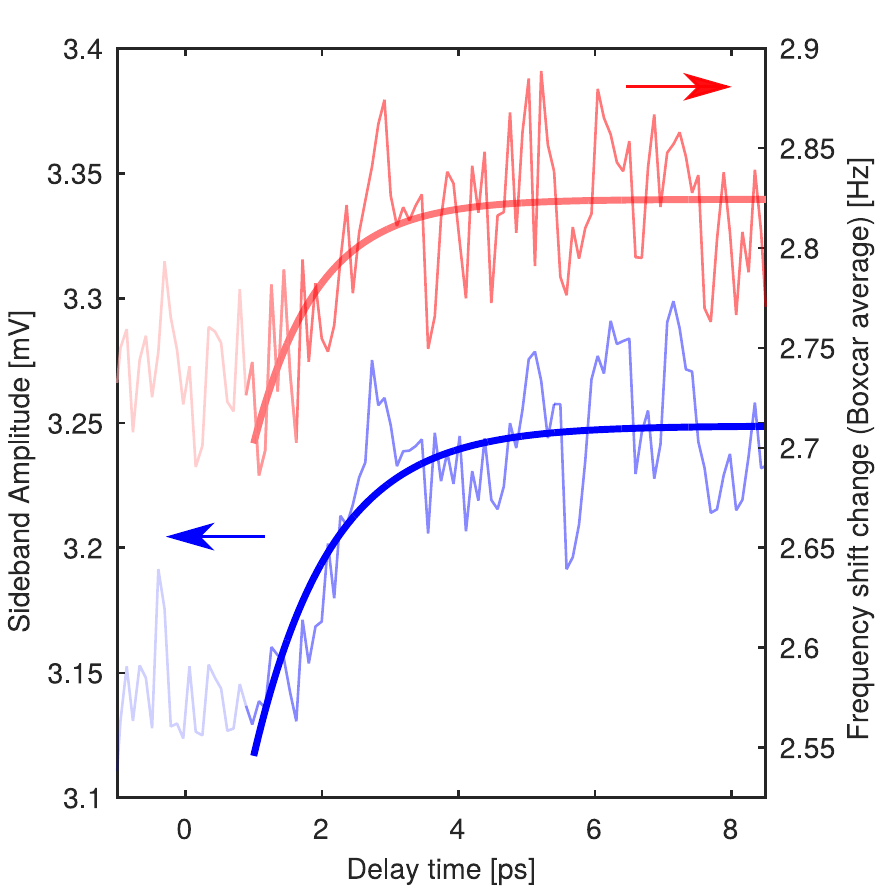}
    \caption{Photocarrier decay measured in LT-GaAs at constant height (1~nm tip lift) and -950~mV bias. A 780~nm pump laser and a 610~nm probe laser is used. Decay time recording using the direct sideband detection (blue) compared to the boxcar averaging in the time domain (red) is shown. The decay is fitted to an exponential decay with a time constant of $\tau=1.1\pm0.4$~ps for sideband and  $\tau =0.9\pm0.6$~ps for boxcar detection)}
    \label{fig:ltgaasUHF}
  \end{figure}

The recorded change in frequency shift at constant bias as a function of the delay time between the two laser pulses is plotted in figure \ref{fig:ltgaasUHF}. The direct sideband detection is performed simultaneously with the boxcar averaging. The data was fitted with a simple exponential decay since a low fluence is used \cite{McIntosh1997,Segschneider2002}. A decay time of $
\tau=1.1\pm0.4$~ps for the sideband detection and  $\tau =0.9\pm0.6$~ps for the time domain 
boxcar averaging is measured by fitting an exponential decay in the form of $ a (1- \exp(-x/\tau)) + c$. This lifetime is in the expected range of 1~ps within error for both measurements. The fitting term  $c$ corresponds to the value at zero delay, hence the first term from equation \ref{eq:avgTrepbaselinecontibution}. The term $a$ is equal to the increase in signal between zero delay and the steady state reached after the exponential decay. The data shown in figure \ref{fig:ltgaasUHF} is recorded in a similar way to the bias curves previously shown. The tip is lifted by 1~nm, the feedback is turned off and a DC-bias is applied. With the feedback off, the delay between the two laser pulses is scanned and the change in frequency shift is recorded.

In general, the sideband detection appears to have a better signal-to-noise ratio due to the use of a lock-in measurement. Additionally, the upper sideband and the lower sideband are added to increase the signal-to-noise ratio. Sideband detection is also favored for imaging applications. On this sample of LT-GaAs we did not find a spatial variation of $\tau$. This is not surprising given the high density of defects in this sample. In general, as the origin of the tip-sample interaction is electrostatic, the best lateral resolution we expect is similar to KPFM and thus at the nm scale \cite{Gross2014}.

\section{Conclusion}

We have derived the fundamental factors determining the fastest possible time resolution of AFM. Surprisingly, time resolution in AFM for a time decaying tip-sample interaction is limited by the thermal noise of the cantilever and not the cantilever mechanical response time or electronic limits of the AFM. We theoretically describe a generic pump-probe scheme with two excitation pulse trains delayed relative to each other that allow for decay time measurements of arbitrary signals and conclude that time resolution is practically limited only by the smallest delay time achievable between the pump and probe pulse. This opens the possibility to measure decay times orders of magnitudes faster than the slow mechanical or electrical characteristic times intrinsic to AFM. We experimentally implement this detection scheme by combining a UHV AFM system with a fs pump-probe laser system to measure the $\sim$1~ps photocarrier lifetime in LT-GaAs. We have thus achieved a time resolution in AFM which is $10^7$ times faster than the oscillation period of the cantilever and $10^{10}$ times faster than the inverse electronic measurement bandwidth. We emphasize that this pump-probe scheme is not limited to optical excitation as long as the response exhibits a decay with time. Repetitive short pump and probe pulse on the order of the decay time and variable delay times smaller than the decay time is all that is needed. Suitable pump-probe excitation schemes can be envisioned for the investigation of the spatial and time resolved dynamics in photoelectric charge generation and recombination, ion mobility or heat transport. Compared to similar techniques such as PIFM, NSOM or STM a major advantage of the presented AFM technique is that it does not need to operate in the near field and is thus less susceptible to spurious artifacts because the measured interaction signal is not exponentially dependent on tip-sample separation.

\section{Acknowledgments}

The authors would like to thank Prof. David Cook and Prof. Carlos Silva for extensive scientific discussion. The authors would like to thank the NSERC and FQRNT for funding. Z.S. would like to thank the SNSF for a DOC.Mobility fellowship. 

\bibliographystyle{ieeetr}
\bibliography{library}
\end{document}